\documentclass{article} % For LaTeX2e
\usepackage{iclr2025_conference,times}

\usepackage{amsmath,amsfonts,bm}

\def\eqref#1{equation~\ref{#1}}
\def\1{\bm{1}}

\DeclareMathAlphabet{\mathsfit}{\encodingdefault}{\sfdefault}{m}{sl}
\SetMathAlphabet{\mathsfit}{bold}{\encodingdefault}{\sfdefault}{bx}{n}

\DeclareMathOperator*{\argmin}{arg\,min}

\usepackage{hyperref}
\usepackage{url}

\usepackage{graphicx}
\usepackage{tikz,lipsum,lmodern}
\usepackage[most]{tcolorbox}
\usepackage{algorithm} 
\usepackage{algpseudocode} 
\usepackage{pifont}

\usepackage{booktabs} % For formal tables
\usepackage{array}    % For table formatting
\usepackage{siunitx}  % For aligning numbers by decimal point
\usepackage{subcaption}
\usepackage{multirow}
\usepackage{color, colortbl}
\definecolor{LightCyan}{rgb}{0.94,1,1}
\definecolor{LightRed}{rgb}{1,0.94,0.94}
\definecolor{Gray}{rgb}{0.8,0.8,0.8}
\definecolor{LightGray}{rgb}{0.92,0.92,0.92}
\definecolor{LLightGray}{rgb}{0.97,0.97,0.97}
\definecolor{mygreen}{rgb}{0.4,0.7,0.306}
\definecolor{myblue}{rgb}{0.294,0.447,0.796}
\definecolor{mediumred}{RGB}{220, 60, 50}
\usepackage{hhline}
\usepackage{nicematrix}
\usepackage{wrapfig}
\usepackage{tabularx}
\usepackage{xcolor}

\usepackage{comment}

\newcommand*\mycircled[1]{\tikz[baseline=(char.base)]{
            \node[shape=circle,fill,inner sep=0.5pt] (char) {\textcolor{white}{#1}};}}

\title{KidSpeak: A General Multi-purpose LLM for Kids' Speech Recognition and Screening}

\author{
\textbf{Rohan Sharma}\thanks{Corresponding author: \texttt{rohanjag@buffalo.edu}} \quad
\textbf{Dancheng Liu} \quad
\textbf{Jingchen Sun} \quad
\textbf{Shijie Zhou} \quad
\textbf{Jiayu Qin} \\[0.8em]
\centerline{\textbf{Jinjun Xiong} \quad \textbf{Changyou Chen}} \\ [0.8em]
\centerline{Department of Computer Science and Engineering, State University of New York at Buffalo}
}

\iclrfinalcopy % Uncomment for camera-ready version, but NOT for submission.
\begin{document}

\maketitle

\begin{abstract}
With the rapid advancement of conversational and diffusion-based AI, there is a growing adoption of AI in educational services, ranging from grading and assessment tools to personalized learning systems that provide targeted support for students. However, this adaptability has yet to fully extend to the domain of children's speech, where existing models often fail due to their reliance on datasets designed for clear, articulate adult speech. Children, particularly those in early developmental stages or with speech and language pathologies, present unique challenges that current AI models and datasets are ill-equipped to handle. To address this, we introduce \textsc{KidSpeak}, a multi-task speech-enhanced Foundation Model capable of both generative and discriminative tasks specifically tailored to children's speech patterns. Our framework employs a two-stage training process that incorporates \textit{phonetic knowledge} into the speech encoder, achieving an average accuracy of 87\% across four separate tasks. Furthermore, recognizing the limitations of scalable human annotation and existing speech alignment tools, we propose the \textsc{Flexible and Automatic Speech Aligner (FASA)} and leverage the method to construct high quality datasets for training and evaluation. This novel alignment tool significantly improves the quality of aligned children's speech from noisy data, enhancing data quality by 13.6× compared to human annotations, as demonstrated on the \textsc{CHILDES} dataset. To the best of our knowledge, \textsc{KidSpeak} and \textsc{FASA} represent the first comprehensive solution designed for speech and language therapy in children, offering both a multi-purpose speech LLM and a robust alignment tool. 
%Code is available at \href{https://anonymous.4open.science/r/children_whisper-9BB0/README.md}{Here}.
\end{abstract}

\section{Introduction}
% \cy{Please talk about the challenges of kids speech data compared to standard speech data.}

Humans begin to acquire the fundamental cues of vocal communication as early as \textbf{3 months of age} \citep{us2017speech}. As development advances, some individuals master their vocal abilities to such a degree that they are capable of vocalizing with over \textbf{1000~$\mathbf{kHz}$}, allowing for the conveyance of complex and nuanced ideas and emotions \cite{garnier2010vocal}\footnote{For reference, the normal adult male voice ranges from 90 to 155 Hz, while adult female voices range from 165 to 255 Hz \citep{fitch1970modal}.}. On the other hand, \textit{hearing} begins as early as the \textbf{28$^{\text{th}}$ week of gestation} in humans \citep{querleu1988fetal}, eventually leading to an auditory capacity capable of discerning frequencies as precise as \textbf{0.5~$\mathbf{Hz}$} within a range of 20 Hz to 20 kHz \citep{romand2014development}. Despite these remarkable developmental milestones, numerous challenges persist in early speech acquisition. In fact, nearly \textbf{1 in 12 children} in the U.S. aged 3 to 17 has experienced a disorder affecting voice, speech, language, or swallowing, with almost half of them not receiving any intervention services in the past year \citep{black2015communication}. These statistics highlight that, despite our advanced auditory and vocal capabilities, we continue to face significant barriers in the early diagnosis and treatment of speech-related disorders. 

\begin{figure}[t!]
  \centering
  \begin{minipage}[b]{0.60\textwidth}
    \includegraphics[width=\textwidth]{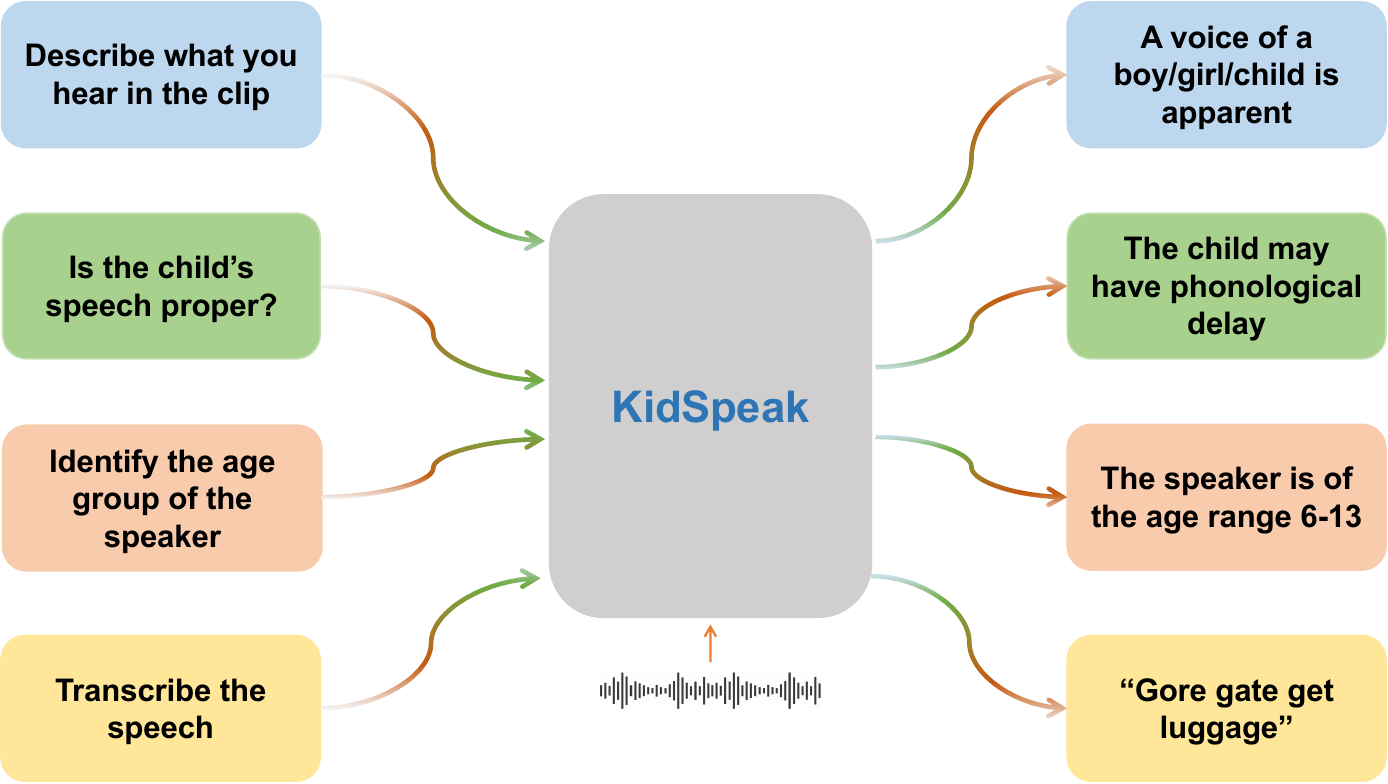}
  \end{minipage}
\caption{\textbf{Overview:} We propose \textsc{KidSpeak}, a multi-purpose speech based LLM aimed at diagnosis and transcription of kids' speech. The framework leverages a customized speech encoding procedure incorporating phonetic information enhancing downstream performance.}
\label{fig: whisper}
\end{figure}

The setbacks are further compounded by the dearth of data pertaining to kids' speech and vocalizations, which poses additional challenges towards the development of automated computational tools. Consequently, the current state-of-the-art ASR systems remain limited to the use of the widely popular datasets of \textsc{Librispeech} \citep{panayotov2015librispeech} and \textsc{Librivox} \citep{shankar2024coraal}, while also incorporating some of the lesser-known open-source datasets such as \textsc{WSJ} \citep{garofolo1993csr}, \textsc{CORAAL} \citep{shankar2024coraal}, and \textsc{TED-LIUM} \citep{rousseau2012ted}, along with their processed versions. Furthermore, the dataset creation phase itself for kids is burdened by the \textit{labour-intensive nature} of the task, considering the nuanced speech and articulation patterns of children. This often leads to current approaches assuming proper pronunciation and articulation, \textbf{failing to account} sufficiently for kids' speech, and faltering \textit{significantly more} with \textbf{accented and non-native kids' speech}, often generating \textit{offensive and inaccurate transcriptions} \citep{ramesh2022beach}.
As an exemplar, the following transcription showcases a child's speech using the state-of-the-art ASR systems, \textsc{Whisper} \citep{radford2023robust} and \textsc{Wave2Vec 2.0} \citep{baevski2020wav2vec20frameworkselfsupervised}. The child is a \textbf{4-year-old non-native boy}.

\begin{tcolorbox}[colback=red!5!white,colframe=red!55!black]
\textbf{Utterance}:\quad\textit{and they are looking at the frog; and because he cracked his egg}

\textbf{\textsc{Whisper}}:\quad \;  \textit{and they recognize the fog; and because do you grab this egg?}

\textbf{\textsc{Wav2Vec}}:\quad \textit{   unfated in that the fog; and because fee practis ed}
\end{tcolorbox}

In an attempt to overcome these major hurdles, we introduce \textsc{KidSpeak}, a speech-based LLM with multi-task capacities of ASR, gender and dialect identification, and speech pathology classification, trained on a curated corpus of kids' speech through instruction tuning. The framework is based upon the foundations of spoken language understanding adapted towards kids' speech transcription and diagnosis of speech language pathologies. We train the method using a specialized two-stage procedure, wherein we utilize simultaneous phonetic and English transcription as a pre-training task for the \textsc{Whisper} ASR model in order to incorporate \textit{phonetically informed encoding} capacities into the encoder of \textsc{Whisper}, as the first stage. Subsequently, the encoder of the model is used in the final framework. 

Additionally, we acknowledge several limitations inherent in the existing datasets for children's speech. Given the \textbf{scarcity of relevant data}, we explore the \textsc{CHILDES} \citep{macwhinney2000childes} corpus as a resource for children's speech. However, it is important to note that the transcriptions within this dataset are \textbf{significantly compromised}. The annotators involved are often engaged in multifaceted tasks, as the children included in the corpus frequently exhibit speech and language disabilities. Consequently, some annotators focus specifically on issues such as stuttering or speech sound disorders, while others address dialectical variations. This diversity in annotation purpose leads to \textbf{inconsistencies in human transcriptions}, limiting their applicability for developing robust automated systems. We therefore develop a new forced alignment tool \textsc{Flexible and Automatic Speech Aligner (FASA)}, allowing us to extract accurate, aligned, and well-segmented audio segments and the corresponding transcriptions under flexible conditions, creating a corpus for \textsc{KidSpeak}. \textbf{Our main contributions} are, \textbf{\mycircled{1}} We develop \textsc{KidSpeak}, a novel multi-task speech-based foundation language model aimed at diagnosis and transcription of children's speech. \textbf{\mycircled{2}} We innovate a two-stage training procedure for the audio encoder, in order to incorporate phonetic information into the encoder, provably enhancing the downstream performance of the framework. \textbf{\mycircled{3}} We develop the \textsc{Flexible and Automatic Speech Aligner}, a novel forced alignment tool, enabling extraction of accurate and aligned audio from noisy speech and demonstrate its utility over the \textsc{CHILDES} corpus in our framework.

\section{Related Work}
Availability of data over the visual and descriptive-visual domain has spurred a preponderance of work towards understanding the visual domain. We witness a similar emergence in the aural understanding domain. Herein we provide an abridged summary of the contemporary work relevant to this manuscript. A more detailed description of the literature is provided in the Section \ref{sec: lit} under the Appendix.

\begin{comment}
\textbf{Visual Understanding:}  
Recent advancements in multi-modal understanding, driven by large datasets, have led to significant progress in visual and linguistic modalities. Works like LLaVA \citep{liu2024visual}, PaLME \citep{driess2023palm}, Flamingo \citep{alayrac2022flamingo}, BLIP \citep{li2022blip}, and GPT-4 \citep{achiam2023gpt} employ instruction-based tuning to integrate visual tokens into language models via self-attention (LLaVA, PaLME), cross-attention (Flamingo), or separate networks (BLIP). These approaches have applications in enhancing search engines, creative tasks, and improving accessibility, such as augmentative communication \citep{chanjaradwichai2019trackable} and sign language recognition systems \citep{gong2024llms}. {\em Our work is a generalization of these frameworks into the audio domain.}
\end{comment}

\textbf{Speech-based LLMs and Spoken Language Understanding:}  
Challenges in encoding speech with LLMs stem from handling long sequences of audio. GSLM \citep{lakhotia2021generative}, TWIST \citep{hassid2024textually}, and SpeechGPT \citep{zhang2023speechgpt} address them by using quantized speech representations using models such as HuBERT \citep{hsu2021hubert}. While others employ log-mel spectrograms to develop representations which are then combined with textual data for multi-modal generative tasks, such as speech recognition, generation and understanding \citep{fathullah2024prompting, nachmani2023spoken, zhao2023librisqa, gong2023listen}, using ASR models such as \textsc{Whisper} \citep{radford2023robust}, \textsc{Wav2Vec} \citep{baevski2020wav2vec20frameworkselfsupervised}, Conformer \citep{gulati2020conformer}, and AST \citep{gong2021ast}, or using multimodal retrieval based models such as ImageBind \citep{girdhar2023imagebind} like the PandaGPT \citep{su2023pandagpt}. {\em Our work innovates techniques essential for processing and understanding of kids' speech.}

\textbf{Kids' Speech:}  
A key limitation of current works is the insufficient handling of nuanced speech variations, such as accents, dialects, intonations, and developmental or disordered speech, as is typical in children. The field of children's speech recognition remains under-researched, with only a few notable approaches, such as LSTM-based disfluency detection \citep{venkatasubramaniam2023end} and teacher-student models \citep{plantinga2019towards}. {\em To the best of our knowledge, this work is the first to propose leveraging LLMs as multi-task models with diagnostic capabilities for children's speech, offering substantial potential in the domain of speech therapy and supporting Speech-Language Pathologists.}

\textbf{Forced-Alignment Toolkits} \label{forced_alignment_related_works} Traditional audio-transcription alignment relies on human annotators \citep{praat,grover2020audino}, which is not scalable for large datasets. While \citet{KISLER2017326} offers components of a forced-alignment pipeline, it does not solve the alignment issue. \citet{GAN2019} uses GANs for data augmentation in children's ASR datasets but does not introduce new data. Several studies utilize human-labeled transcriptions for forced-alignment \citep{mcauliffe17_interspeech, POnSS, zhang2023speechgpt,Liu_MacWhinney_Fromm_Lanzi_2023}, with the Montreal Forced Aligner (MFA) being prominent \citep{mcauliffe17_interspeech}. However, MFA demands perfect alignment, limiting its effectiveness. {\em Our work improves over existing works and over human annotators by margins of over $13\times$.}

\section{Method}\label{sec: method}

% \cy{Please first give an overview of the method}

\begin{figure}[ht]
    \centering
    
    \includegraphics[width=0.7\textwidth, height=1.7in]{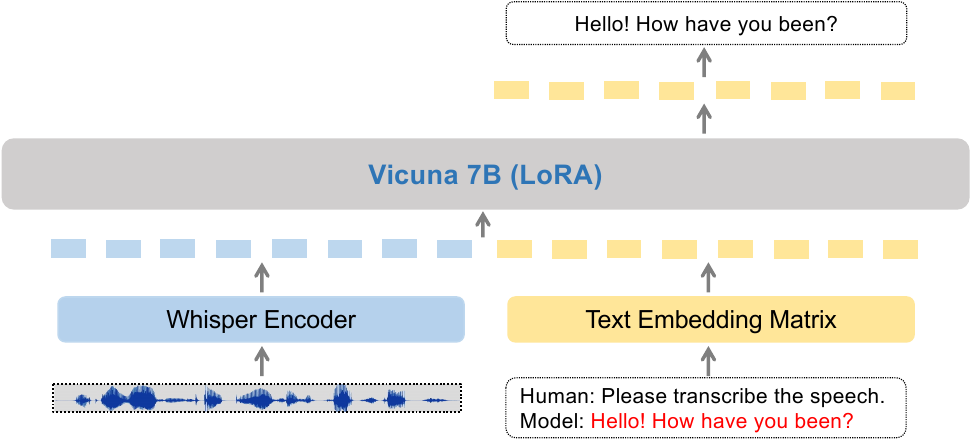}  
    
    %\vspace{0.2cm} 
    
    %\includegraphics[width=0.7\textwidth, height=.8in]{stacking.pdf}
    
    \caption{\textbf{The Proposed Framework:} \textsc{KidSpeak} uses phonetically informed speech features from the pre-trained multi-head \textsc{Whisper} encoder. The features are concatenated with the text embeddings of the instructions during training endowing the framework with spoken context and textual instruction through self-attention.}
    \label{fig: framework}
\end{figure}

We describe the method that we implement in order to create a multi-purpose speech LLM that exhibits potential as a useful diagnosis tool for speech-related impairments. The framework is trained using instruction finetuning. In summary, we utilize an audio encoder pre-trained using a targeted procedure, to generate representations which are subsequently post-processed and prepended to the textual embeddings and processed by a pre-trained LLM in order to generate answers. The main framework is illustrated in Figure \ref{fig: framework}. We employ the pre-trained Vicuna 7B model \citep{vicuna2023} as our main LLM\footnote{We choose Vicuna as it is one of the best LLMs we began with. Due to computational contraints, we did not try other strong pretrained LLMs. However, we believe they would perform similarly and our conclusions remain.}. We note that our data exhibits a significant domain gap from the pre-training, both in terms of format (audio vs text) and content (kids' speech). Therefore, in order to retain the general capacity of the LLM and avoid overfitting to the newer data, we finetune the LLM using Low Rank Approximation (LoRA) \citep{hu2021lora}. This additionally helps with the memory footprint of the model, allowing for larger batch sizes.

\begin{wrapfigure}{R}{0.4\textwidth}
    \begin{minipage}[t]{.93\linewidth}
        \vspace{-.1in}
        \centering
        \includegraphics[width=\linewidth]{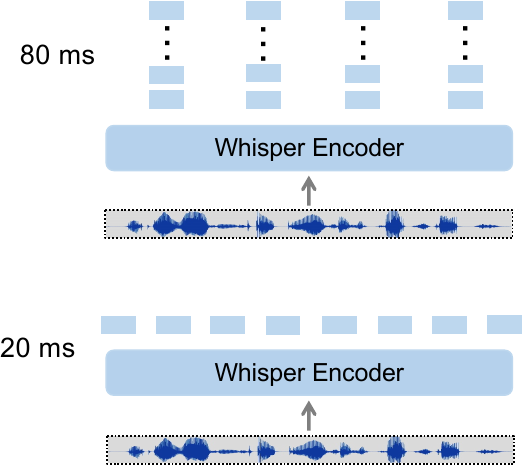} 
        \caption{The speech embeddings are post-processed through a stacking mechanism (Top), ensuring adequate granularity. Thereafter the stacked features are projected onto the feature space of the LLM, ensuring synchronization between the two modalities.}
        \label{fig: stacking}
    \end{minipage}
\end{wrapfigure}

\subsection{Speech Based LLM}
\begin{figure}[h]
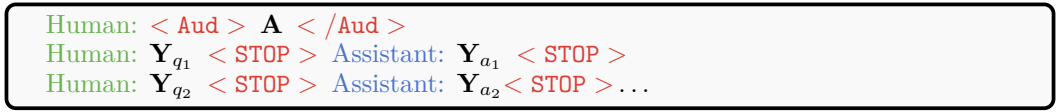

    \centering
\begin{tcolorbox}[colback=gray!5!white,colframe=black!55!black]
\vspace{-.08in}
    $\textcolor{mygreen}{\text{Human: }} \textcolor{mediumred}{<\mathtt{Aud}>} \hspace{0.5em} \mathbf{A} \hspace{0.5em}\textcolor{mediumred}{</\mathtt{Aud}>} \\ \textcolor{mygreen}{\text{Human: }} \mathbf{Y}_{q_1} \hspace{0.5em} \textcolor{mediumred}{\mathtt{<STOP>}} \hspace{0.5em} \textcolor{myblue}{\text{Assistant: }}  \mathbf{Y}_{a_1} \hspace{0.5em} \textcolor{mediumred}{\mathtt{<STOP>}}\\ \textcolor{mygreen}{\text{Human: }} \mathbf{Y}_{q_2} \hspace{0.5em} \textcolor{mediumred}{\mathtt{<STOP>}}  \hspace{0.5em} \textcolor{myblue}{\text{Assistant: }} \mathbf{Y}_{a_2} \textcolor{mediumred}{\mathtt{<STOP>}} \ldots$
\vspace{-.08in}
\end{tcolorbox}
\caption{\textbf{Instruction Template: }We illustrate two instructions in the general input sequence that we implement to for the IFT procedure. The conversation structure comprises alternating exchanges between a human user and the \textsc{KidSpeak}, where tags $\textcolor{mediumred}{<\mathtt{Aud}>}$ and $\textcolor{mediumred}{</\mathtt{Aud}>}$ demarcate the audio representations. The framework is trained to predict $\mathbf{Y}_{a_t}$ using the aural and instructional context. The $\textcolor{mediumred}{\mathtt{<STOP>}}$ is set to \texttt{\#\#\#} in practice.}
    \label{fig: instruction_template}
\end{figure}
We employ a \textsc{Whisper}-based encoder for speech in our main framework. The audio representations from \textsc{Whisper} are prepended to the text embeddings consisting of instructions and the teacher-forced output. The complete sequence is further processed using the Vicuna LLM, incorporating self-attentive mechanism thereby incorporating audio-lingual context in order to learn and generate informed inference. However, we note that the native implementation of the \textsc{Whisper} encoder generates encodings of shape $batch\_size\times1500\times768$ leading to a significant increase in the memory footprint of the model due to the extensive sequence length and the consequent self-attention matrices. We therefore apply a post-processing step to the feature tensors produced by the \textsc{Whisper} encoder. This step aims to reduce the final input sequence length whilst minimizing information loss. In this procedure, we aggregate multiple consecutive audio features (Figure \ref{fig: stacking}) to form a cumulative representation that spans 80 milliseconds per feature vector. The aggregated features are then processed through a two-layer adapter network to align the feature space dimensions with those of the textual embeddings, which are further processed using the LLM. For each audio sample $i$, we create a multi-turn instruction following dataset $(\mathbf{Y}^{(i)}_{q_1}, \mathbf{Y}^{(i)}_{a_1} ... \mathbf{Y}^{(i)}_{q_T}, \mathbf{Y}^{(i)}_{a_T})$ illustrated in the Figure \ref{fig: instruction_template}, wherein the instructions are randomly ordered during training. The framework is then trained using a conditional auto-regressive prediction objective 
\begin{equation}
    \argmin_{\theta_{s}, \theta_{m}} 
\sum_{i=1}^{N} \sum_{j=1}^{T} \mathcal{L}^{(i)}_{j}, 
\quad \text{with} \quad
\mathcal{L}^{(i)}_{j} = 
- \sum_{t=1}^{T^{(i)}_{a_j}} 
\log P\left(
y_{a_{j, t}}^{(i)} \mid 
y_{a_{j, <t}}^{(i)}, 
\mathbf{A}^{(i)}, 
\mathbf{Y}_{q_j}^{(i)}; 
\{\theta_{enc}, \theta_{s}, \theta_{m}\}
\right),
\end{equation} %\label{eqn: ar_main}
for a sample indexed $i$ and the instruction $j$. The speech contexted LoRA parameters $\theta_s$ and the adapter MLP parameters $\theta_m$ are estimated, conditioning upon the speech sample $\mathbf{A}^{(i)}$ represented due to the frozen speech encoder parameters $\theta_{enc}$ and the instruction $\mathbf{Y}_{qj}$. We train the framework to predict the answer $\mathbf{Y}^{(i)}_{a_j}$ of length $T^{(i)}_{a_j}$. Additionally, we find that the a targeted encoding scheme for $\theta_{enc}$ benefits the framework, as we detail next.

\subsection{Multi-head Whisper Pretraining} \label{sec: mh_method}

A preponderance of developing speech is characterized by phonetic challenges wherein the child mispronounces similar phonetic units \citep{munson2012role}. Speech and language therapists must proficiently utilize phonetics for transcription to accurately diagnose and treat speech-sound disorders \citep{munson2012role, ball2002transcribing}. Therefore, reliable pre-training in  phonetic transcription is imperative, as inaccuracies can significantly affect clinical management and therapeutic outcomes. In recognition of these facets in diagnosis, we conduct a separate procedure in order to endow the speech encoder with phonetic information. The audio encoder employed is based on the \textsc{Whisper} model which is designed to encode mono-channel audio sampled at 16 kHz, which is then transformed into log-Mel spectrogram images. The encoder generates audio features, with each feature tensor corresponding to a 20-millisecond segment of audio. For optimal performance, we utilize the default configuration, which processes audio in 30-second chunks, and affix it with separate dedicated decoders for the orthographic English transcription and phonetic transcription as illustrated in Figure \ref{fig: whisper}. This essentially yields a training regime wherein the same encoder facilitates both textual and phonetic transcriptions while the two decoders specialize in decoding the features into phonetic and English transcriptions respectively. The model is then trained using the next token prediction loss for both decodings simultaneously. The objective is,
\begin{align} \label{eqn: ar_whisper}
    \argmin_{\theta_{enc}, \theta_{dec}^{(en)}, \theta_{dec}^{(ph)}} & \sum_{i=1}^{N} \mathcal{L}^{(i)}_{ar}, \quad \text{where} \\  \nonumber
    \mathcal{L}^{(i)}_{ar} = - \sum_{t=1}^{T_e^{(i)}} \log P\left(y_{t, e}^{(i)} \mid y_{<t, e}^{(i)}, \mathbf{A}^{(i)}; \{\theta_{enc}, \theta_{dec}^{(en)} \}\right) &- \sum_{t=1}^{T_p^{(i)}} \log P\left(y_{t, p}^{(i)} \mid y_{<t, p}^{(i)}, \mathbf{A}^{(i)}; \{\theta_{enc}, \theta_{dec}^{(ph)}\} \right),
\end{align}

where $\mathcal{L}^{(i)}_{ar}$ is the combined auto-regressive loss for the $i$-th sample, $T_e^{(i)}$ and $T_p^{(i)}$ are the lengths of English and phonetic transcriptions, $y_{t, e}^{(i)}$ and $y_{t, p}^{(i)}$ are the target tokens, $y_{<t, e}^{(i)}$ and $y_{<t, p}^{(i)}$ are preceding tokens and $\mathbf{A}^{(i)}$ is the input audio. We specify the decoder parameters separately with $\theta_{dec}^{(en)}$ and $\theta_{dec}^{(ph)}$ representing the English and the phonetic decoders respectively using the same encoder parameters $\theta_{enc}$. We also utilize a special token $|\mathrm{phn}|$ to signify phonetic decoding. However, we hypothesize that explicit measures towards alignment of the two decoders may lead to enhanced generalization and understanding of the language through a unified feature space representation. Therefore, in order to enhance the alignment of the transcription mechanisms, we further process the downstream features of the decoders using two additional mechanisms.

\begin{figure}[!h]
  \centering
  \begin{minipage}[b]{0.70\textwidth}
    \includegraphics[width=\textwidth]{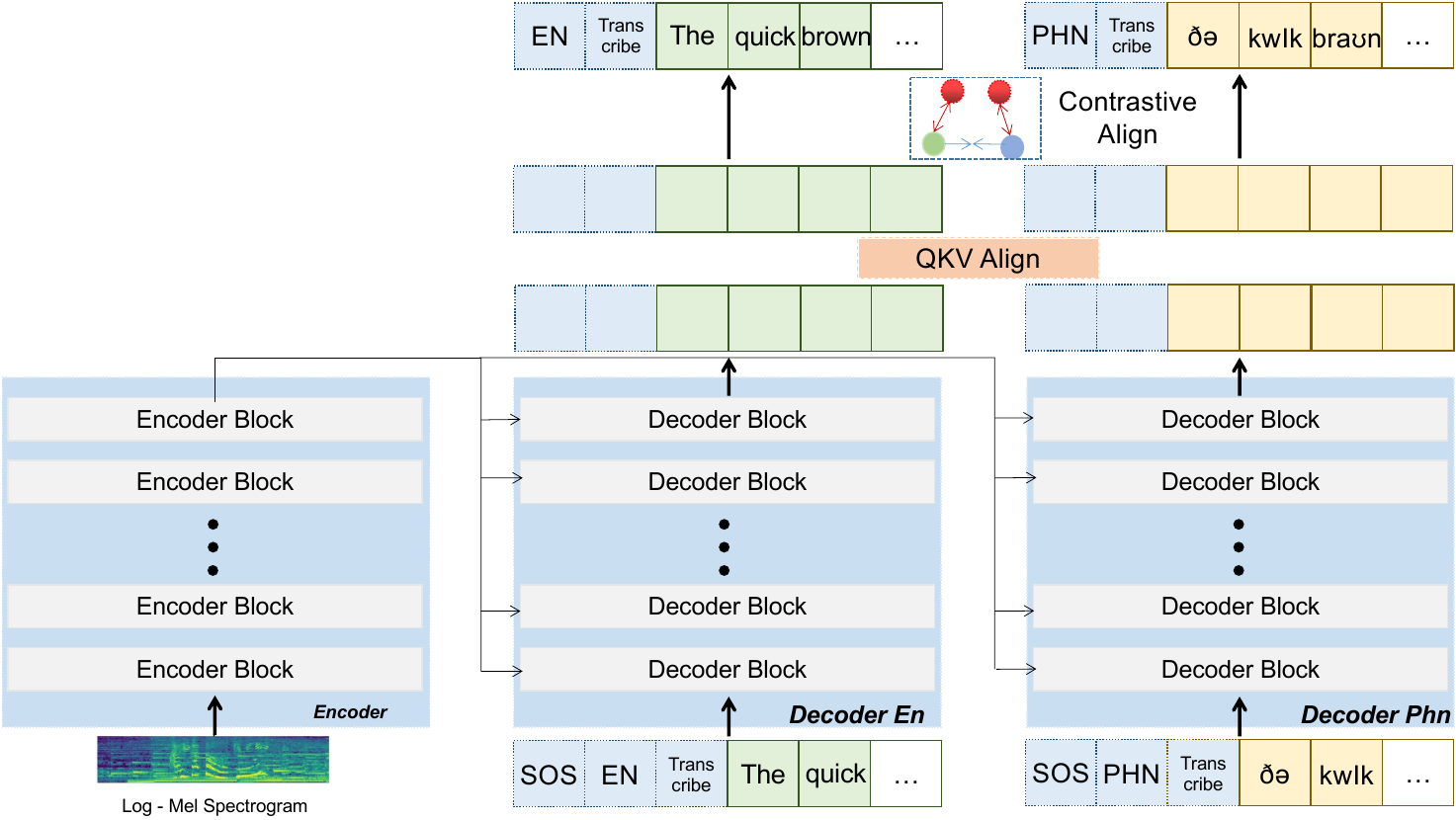}
  \end{minipage}
\caption{\textbf{Multi-head \textsc{Whisper}:} We employ two separate decoders to decode the same speech segment in English and its Phonetic counterpart. The decoders are further aligned using contrastive and cross-attentive mechanisms, synchronizing the procedure.}
\label{fig: whisper}
\end{figure}

\paragraph{Contrastive Alignment}
A cross-entropy-based contrastive loss \citep{https://doi.org/10.48550/arxiv.2002.05709} is incorporated, utilizing the $|\mathrm{startoftranscript}|$ token from both decoders to align pairs of English and phonetic sequences derived from the same audio more closely, while simultaneously separating sequences originating from different audio samples, formulated as,

\begin{equation}
\mathcal{L}_{\text{con}}^{(i)} = - \log \frac{\exp \left( \text{sim}\left(\mathbf{h}_{\text{sos}, e}^{(i)}, \mathbf{h}_{\text{sos}, p}^{(i)}\right) / \tau \right)}{\sum_{j=1}^{B} \exp \left( \text{sim}\left(\mathbf{h}_{\text{sos}, e}^{(i)}, \mathbf{h}_{\text{sos}, p}^{(j)}\right) / \tau \right)},
\end{equation}
where $\mathcal{L}_{\text{contrastive}}^{(i)}$ represents the contrastive loss for the $i$-th sample, where $\mathbf{h}_{\text{sos}, e}^{(i)}$ and $\mathbf{h}_{\text{sos}, p}^{(i)}$ are the hidden states for the English and phonetic decoders, respectively, $\text{sim}(\cdot, \cdot)$ denotes the similarity function, $\tau$ is the temperature parameter, and the denominator sums over similarities over a batch of size $B$.

\paragraph{Cross-attentive Alignment}

This mechanism leverages cross-attention to synchronize the hidden states from two decoders derived from the same audio input.  We implement the mechanism over the final hidden states of both decoders as formulated below, 

\vspace{-.12in}
\begin{align}
\text{Residual}_p^{(i)} &= \left( \text{Softmax} \left( \frac{(\mathbf{H}_e^{(i)} U) (\mathbf{H}_p^{(i)} U)^\top}{\sqrt{d_k}} \right) (\mathbf{H}_p^{(i)} U) \cdot D \right) + \mathbf{H}_p^{(i)}, \\
\text{Residual}_e^{(i)} &= \left( \text{Softmax} \left( \frac{(\mathbf{H}_p^{(i)} U) (\mathbf{H}_e^{(i)} U)^\top}{\sqrt{d_k}} \right) (\mathbf{H}_e^{(i)} U) \cdot D \right) + \mathbf{H}_e^{(i)},
\end{align}
\vspace{-.1in}

where $ \mathbf{H}_e^{(i)} $ and $ \mathbf{H}_p^{(i)} $ represent the hidden states of English and phonetic transcriptions, $ U (768\times1024)$ and $ D (1024\times768)$ the upward and downward projection matrices, and $ d_k $ the dimensionality of the keys. The upward projection enriches the representation space to capture detailed alignments, while the downward projection ensures the contextualized outputs are compatible with the original dimensions. In practice, we implement this mechanism using multi-head attention with 16 heads. The residuals are ultimately used in the evaluation of $\mathcal{L}_{ar}$ in Equation \ref{eqn: ar_whisper}. A detailed justification for this scheme is provided under Section \ref{sec: mh_justification}. In summary, these alignment mechanisms ensure that phonetic and orthographic transcriptions are not only aligned but also mutually reinforcing, enhancing the encoder's ability to generate accurate and contextually relevant transcriptions for both modalities, ultimately improving the model's overall performance and utility in downstream tasks. The model is subsequently trained end-to-end using a linear combination of the two resulting loss functions given by $\mathcal{L}_{whisper}^{(i)} = \mathcal{L}_{ar}^{(i)} + \lambda \mathcal{L}_{con}^{(i)}$.

\newcommand{\mname}{FASA}
\section{Dataset Construction}

In this section, we present the \textsc{Flexible and Automatic Speech Aligner (FASA)}, a novel toolkit designed for forced alignment to create high-quality fine-tuning datasets. A robust children's speech model necessitates a large, diverse dataset with accurately aligned audio and transcriptions. However, obtaining such high-quality data is challenging due to the distinctive speech patterns of children, particularly those with speech and language disorders. Human annotation is labor-intensive and requires domain expertise, as noted by \citet{Miller_Andriacchi_Nockerts_2016}, with our experience showing that annotating a single audio segment can take \textit{3-8 times longer} than its duration. Additionally, the quality of annotations varies significantly, especially in datasets like \textsc{CHILDES} \citep{CHILDES}, which cater to diverse transcription purposes, resulting in many transcriptions being incomplete or irrelevant. To enhance \textsc{KidSpeak}, a general-purpose forced alignment toolkit is crucial for extracting high-quality children's speech datasets from low-quality sources. Existing methods, such as \textsc{MFA} \citep{mcauliffe17_interspeech}, rely on accurate transcriptions, which are often impractical to obtain. Therefore, we propose a flexible and automated forced alignment toolkit that addresses various challenges in current children's speech datasets.

\subsection{FASA Design}

Given a non-timestamped audio file and its noisy or incomplete transcription, forced alignment generates time-stamped audio segments paired with high-quality transcriptions. \textsc{KidSpeak}, like many modern automatic speech recognition systems, requires input audio to be divided into smaller segments during training. For instance, the \textsc{Whisper} model \citep{radford2022whisper} pads or trims audio inputs to 30 seconds. Consequently, when associating a non-timestamped transcription with a lengthy audio file, a forced-alignment toolkit is essential for creating a model-compatible dataset. Formally, the forced-alignment task involves an audio sample containing \( n \) utterances, \( \mathbf{A}:=\{A_1,A_2,...A_n\} \), and a transcription of \( m \) “words,” \( T:=\{T_1,T_2,...T_m\} \). A “word” in \( T \) represents a fundamental unit of transcription, which may refer to a sentence, a single word, or a phonetic symbol. The goal is to associate each \( A_i \) with its corresponding words in \( T \), from \( T_{si} \) to \( T_{ei} \), or indicate that \( A_i \) lacks a transcription in \( T \). We denote this association as \( A_i = (T_{si}, T_{ei}) \).
A robust auto-alignment system should exhibit two crucial features. First, it must not assume that if \( A_i= (T_{si}, T_{ei}) \) and \( A_j= (T_{sj}, T_{ej}) \) with \( i<j \), then \( ei<sj \); an utterance appearing earlier in the audio does not guarantee its early appearance in the transcription. Second, \( A_i \) may lack a corresponding \( (T_{si}, T_{ei}) \), implying that \( A_i=\emptyset \). This means not all audio segments have transcriptions, and some audio may remain untranscribed. Similarly, \( T_k \in T \) does not imply \( T_k \in A \); not every word in the transcription corresponds to an audio segment. These features are essential as they relax the need for completeness and order in the provided transcription, mirroring more realistic scenarios. While a common method for obtaining large datasets of paired audio and transcriptions is through Internet scraping, many online transcriptions are noisy and incomplete, often with missing or misordered entries. Under these conditions, existing forced-alignment toolkits, such as those proposed by \citet{mcauliffe17_interspeech}, are inadequate. Further details are discussed in Appendix \ref{forced_alignment_related_works}.

\begin{figure}[t]
  \centering
 \includegraphics[width=0.70\textwidth]{FASA_paper.png}
  \caption{\textbf{Pipeline of FASA:} The input is an audio file and a transcription. Module \textcircled{1} optionally cleans the input transcription; module \textcircled{2} segments and makes predictions on the audio; module \textcircled{3} forced-aligns audio segments with the provided transcription using Algorithm 1; module \textcircled{4} performs post-generation checking (PGC); and module \textcircled{5} allows user to augment dataset via manual selections. 
  The entire system besides module \textcircled{5} is automatic.}
  \label{fig:pipe}
  \vspace{-7mm}
\end{figure}

\begin{equation}
GT_{k}=
\begin{cases}
  T_{A_k}=\{T_i,T_{i+1},...,T_j\}, & \text{if}\ T_{A_k} \in T \\
  \emptyset, & \text{otherwise}
\end{cases}
\label{eq}
\end{equation}

\subsection{Workflow} \label{sec: fasa_workflow}
\mname~ follows a five-module pipeline to automatically segment, label, and align a long audio file with its transcription, as illustrated in Figure \ref{fig:pipe}. Among the five modules, the second and third are mandatory, whereas the other three are optional for enhancing the quality and quantity of the dataset. These five modules together maximize the correctness of forced alignment under flexible conditions. $\textbf{\textcircled{1}}$ The first module applies a regular expression to clean the provided transcriptions and to exclude any non-alphanumeric characters. $\textbf{\textcircled{2}}$ For the second module, modern ASR models will be used to obtain word-level timestamps of the transcriptions. 
Currently, sentence-level separations from the provided model are used as the segmentation marks for long audio. 
$\textbf{\textcircled{3}}$ After the second module, a folder consisting of audio segments and their corresponding predictions will be generated. The set of predictions for sentence-level utterances will be denoted as $\bar{T} = \{\bar{T_1}, \bar{T_2},...,\bar{T_n}\}$. For each utterance $A_k$, its predicted transcription will be $A_k=\bar{T_{A_k}}$. The third module will apply a sliding-window Algorithm \ref{algo:algo} (in Appendix \ref{fasa_pseudocode}) to find the best matching from the provided transcription ($T$) for each utterance ($A_k$). After this module, two datasets will be generated. The first dataset $\text{DATA}_{align}$ is what the algorithm finds close alignment between the prediction and provided transcription that is within a threshold. 
The second dataset $\text{DATA}_{verify}$ is what the algorithm finds slight mismatches between the prediction and the transcription. For $\text{DATA}_{align}$, the provided transcription from $T$ will be used as the ground truth of the utterance. $\textbf{\textcircled{4}}$ The fourth module, post-generation checking (PGC), is an optional module that iterates through $\text{DATA}_{align}$ to find if there are significant mismatches between a second-round prediction and the aligned transcription on sentence length. The implemented metric for PGC is based on the difference in sentence length between the results of a second-round prediction and the aligned transcription. If the difference is greater than a threshold, the utterance and its transcription will be removed from $\text{DATA}_{align}$. $\textbf{\textcircled{5}}$ The fifth module, user selection, is an optional module that launches a graphical-user-interface (GUI) that allows the user to listen to, select, or input correct transcription for each utterance in $\text{DATA}_{verify}$ so that they could be added to the dataset. After the two optional modules, \mname~ assumes the validity of $\text{DATA}_{align}$, which will be used as the final output dataset. Furthermore, \textsc{FASA} features additional qualitative and user-friendly traits as described in Section \ref{sec: fasa_features}.

\begin{table}[h]
    \centering
    \caption{\textbf{Dataset and attribute specifications:} We present the attributes that we avail from the various dataset compiled to create the training corpus. \textbf{Dis.}: Disorder Labels, \textbf{Age}: Exact Ages of the speakers, \textbf{Gen.}: Gender attribute information, \textbf{Trans.}: Speech transcription for the audio.}
    \label{table:dataset}
    \small
    \begin{tabular}{l|c|c|c|c|c|c|c}
    \toprule
         \textbf{Dataset} &  \textbf{Speakers} & \textbf{Utterances} & \textbf{Time} & \textbf{Dis.} & \textbf{Age} & \textbf{Gen.}& \textbf{Trans.} \\ 
         \midrule
         \rowcolor{LLightGray}UPX & 20 & 2789 & 07:01:23 & \ding{51} & \ding{51} & \ding{51} & -- \\
         \rowcolor{LightGray}CSR & 11 & 639 & 00:26:58 & -- & -- & \ding{51} & \ding{51} \\
         \rowcolor{LLightGray}ENNI (\textsc{FASA}) & 352 & 4402 & 15:16:51 & -- & \ding{51} & \ding{51} & \ding{51}\\ 
         \rowcolor{LightGray}Clinical Eng (\textsc{FASA}) & 1540 & 59539 & 30:11:40 & -- & \ding{51} & \ding{51} & \ding{51} \\ 
         \rowcolor{LLightGray}Clinical Other (\textsc{FASA}) & 292 & 59539 & 4:18:16 & -- & \ding{51} & \ding{51} & \ding{51}\\ 
    \bottomrule
    \end{tabular}
\end{table}

\subsection{Dataset Specifications}
We compile multiple open-source datasets in order to test our framework against a wide variety of instructions, in addition to datasets generated by \textsc{FASA} leading to \textbf{over 57 hours of high-quality data}. The corpus was built in order to adequately represent children with speech pathologies and those with clear speech, and containing a rich collection of speaker related attributes that we aim our method to predict and generate. We summarize the data in Table \ref{table:dataset}. A broader description of the datasets is provided in Section \ref{sec: datasets}.

\section{Evaluation}
We conduct training and performance evaluation of our method across several distinct tasks, each based on specific speech traits. These traits are extracted from the corresponding ground-truth labels available in the datasets referenced in Table \ref{table:dataset}. The tasks are described as follows:

\renewcommand\labelitemi{\ding{117}}
\begin{itemize} 
    \item \textit{Disorder Classification}: Speech disorders are categorized into the following classes: \textbf{\mycircled{1}} inconsistent phonological disorder, \textbf{\mycircled{2}} consistent phonological disorder, \textbf{\mycircled{3}} childhood apraxia of speech, \textbf{\mycircled{4}} phonological delay, \textbf{\mycircled{5}} vowel disorder, \textbf{\mycircled{6}} articulation disorder, and \textbf{\mycircled{7}} no disorder, based on the ground truths provided in the Ultraphonix (UPX) subset of the Ultrasuite repository \cite{eshky2019ultrasuite}. Detailed descriptions of these disorders are provided in Section \ref{sec: disorders_defn}.
    
    \item \textit{Gender Classification}: We conduct binary gender classification using the ground-truth gender labels available in the datasets. Predictions are made only where gender information is explicitly provided.

    \item \textit{Age Group Classification}: Age labels are sourced from the ENNI dataset. To account for the minimal acoustic differences between closely aged children, we divide the age range into two groups: 1–5 years and 6–13 years.

    \item \textit{Transcription}: We compile the transcriptions available for the kids with no speech-related disorders in order to create a reliable benchmark for training and evaluating speech recognition models. We train the framework to transcribe the speech and evaluate using the word error rate and character error rate metrics.
\end{itemize}

We evaluate the performance of the classification tasks using the accuracy of inference. Additional details for the configuration of the training setup are provided under Table \ref{tab: config}.

\paragraph{Phonetic Pre-training Enhances Speech Diagnosis and Transcription in Kids}

\begin{wrapfigure}{R}{0.4\textwidth}
\begin{minipage}[t]{.93\linewidth}
\begin{table}[H]
\vspace{-.35in}

\caption{\textbf{\textsc{Whisper} MH comparison}. We evaluate the Phonetic Error Rate (PER) for our scheme comparing against finetuned open-source ASR models and ablate the alignment objective over the TIMIT dataset \citep{garofolo1993timit}. (MH-1: \textsc{Whisper} with two decoders; MH-2: MH-1 + Cross Attn.; MH-3: MH-2 + Contrastive Alignment)}
\fontsize{6}{7}\selectfont
%\tiny
%\cy{Shoud include the MoCo v3 results from the github, which uses a much larger batchsize}}
\centering
\resizebox{0.99\textwidth}{!}{%
\label{table: whisper_mh}
% \begin{tabular}{@{}l|c|c|c@{}}
\begin{tabular}{l|c}
\toprule
Method      & PER  \\ \midrule
\rowcolor{LLightGray}\textsc{Whisper}  & 10.1    \\
\rowcolor{LightGray}\textsc{Wav2Vec 2.0}  & 9.7   \\
\rowcolor{LLightGray}\textsc{Whisper} MH-1 & 9.6  \\
\rowcolor{LightGray}\textsc{Whisper} MH-2 & 9.2  \\
\midrule
% \rowcolor[HTML]{D9D9D9} 
\rowcolor{LightCyan} 
\textsc{Whisper} MH-3 & \textbf{8.6}  \\
\bottomrule
\end{tabular}
}
%\vspace{-0.3in}
\end{table}
\end{minipage}
\end{wrapfigure}
In addition to garnering benefits over the \textsc{KidSpeak} framework, the aligned training procedure for the \textsc{Whisper} model is beneficial for the transcription performance of the \textsc{Whisper} model as shown in Table \ref{table: whisper_mh}, where we evaluate the Phonetic Error Rate of the models using various configurations. The multi-task evaluation in Table \ref{tab:performance_comparison} compares the performance of \textsc{KidSpeak} and \textsc{KidSpeak} (MH-\textsc{Whisper}) in addition to the PandaGPT \citep{su2023pandagpt}, which we adapt to our application of children's speech diagnosis. While \textsc{KidSpeak} achieves a strong performance in gender classification and disorder classification, the MH-\textsc{Whisper} variant shows notable improvements in disorder classification, word transcription, and character transcription accuracy. Additionally, both methods maintain high accuracy in age-group classification significantly overcoming the performance of PandaGPT. Overall, the MH-\textsc{Whisper} demonstrates a significant enhancement in transcription and classification tasks compared to the original \textsc{KidSpeak}. The performance of the MH-\textsc{Whisper} model, which integrates phonetic and English data, underscores the critical role of phonetic knowledge in tasks related to children’s speech. Children often exhibit unique speech patterns, including phonological disorders and developmental variances. By incorporating phonetic information, the model better understands these nuances, allowing it to differentiate subtle pronunciation variations common among young speakers. This phonetic grounding enhances the model’s ability to generalize across diverse dialects and individual speech patterns, ultimately contributing to more reliable assessments and interventions in speech-related applications for children.

\renewcommand{\arraystretch}{1.0}
\begin{table}[h] 
    \centering
    \caption{\textbf{Multi-Task Evaluation:} We present the evaluations through mean scores over three separate runs for the tasks. $a_{\pm b}$ notation represents mean and standard error of the runs. \textbf{GCA:} Gender Classification Acc., \textbf{DCA:} Disorder Classification Acc., \textbf{WTA:} Word Transcription Acc., \textbf{CTA:} Character Transcription Acc., \textbf{ACA:} Age-group Classification Acc. (\textbf{WTA} = 1 - Word Error Rate and \textbf{CTA} = 1 - Character Error Rate).}
    \label{tab:performance_comparison}
    \small
    \begin{tabular}{ccccccc}
        \toprule
        \textbf{Method} & \textbf{GCA} & \textbf{DCA} & \textbf{WTA} & \textbf{CTA} & \textbf{ACA} & \textbf{Average} \\ 
        \midrule
    \rowcolor{LightGray}\textsc{PandaGPT} & $61.0_{\pm0.3}$ & $42.6_{\pm4.1}$ & $6.6_{\pm1.9}$ & $13.6_{\pm1.2}$ & $84.3_{\pm0.4}$ & $50.3_{\pm2.3}$ \\
    \rowcolor{LightCyan}\textsc{KidSpeak} & $73.8_{\pm0.4}$ & $85.0_{\pm2.9}$ & $82.2_{\pm0.4}$ & $87.0_{\pm0.3}$ & $93.8_{\pm0.5}$ & $84.4_{\pm0.9}$ \\
     \rowcolor{LightCyan}\textsc{KidSpeak} (MH-\textsc{Whisper}) & $73.3_{\pm0.1}$ & $88.8_{\pm2.3}$ & $87.8_{\pm0.2}$ & $91.0_{\pm0.2}$ & $94.1_{\pm0.1}$ & $\textbf{87.0}_{\pm0.6}$ \\
        \bottomrule
    \end{tabular}\label{table: main}
\end{table} 

\paragraph{FASA alignment Outperforms Human Annotators}

Considering the vast size of the dataset, we randomly selected two audio files and transcriptions from the 352 recordings in the ENNI dataset, and report the manual inspection results for data generated by \textsc{FASA} with these files in Table \ref{table:eval}. Several results are worthy of emphasizing here. First, since the transcriptions are noisy, MFA \citep{mcauliffe17_interspeech} completely fails to properly align the audio segments with the correct transcription. \footnote{This is not to say that MFA is not a good model. MFA works fine with high-quality transcriptions that are an approximate match of the audio. However, if the audio/transcription match before the alignment is not good, MFA will not produce anything meaningful.} To be specific, both documents have missing transcriptions corresponding to the beginning of the audio, which results in 99.93\% AW Error. This is because MFA tries to align all the words from the beginning, but since those words do not have available transcriptions, the entire system fails. Second, \textsc{FASA} incorrectly aligns one utterance with its transcription. For that utterance, it misses the ``so the'' sound at the end of the utterance, and the two words are not recorded into the aligned transcription. Manual inspection finds that the speaker stuttered and repeated ``so the", which might be the issue of the model not picking up that trailing sound in the segmented utterance. Lastly, \textsc{FASA}'s result is potentially much better than human annotators.
\citet{attia2023kidwhisper} reports that 5 out of 393 hours of speech in MyST dataset \citep{pradhan2023science} are potentially incorrect with WER$>50\%$, resulting in \textbf{3\%} increase in WER for the entire training dataset. Compared to human annotators that were used to annotate MyST, \textsc{FASA} achieves one magnitude lower WER (\textbf{13.6$\times$}) without requiring any human labor.

\section{Conclusion}
In conclusion, this work presents significant advancements in the field of children's speech analysis. First, we introduce \textsc{KidSpeak}, a pioneering multi-task speech-based foundation language model designed specifically for diagnostic tasks pertaining to children's speech. Second, we propose an innovative two-stage training procedure for the audio encoder that effectively integrates phonetic information, leading to marked improvements in both diagnostic and transcription performance, as well as downstream performances when integrated with \textsc{KidSpeak}. Finally, we develop the \textsc{Flexible and Automatic Speech Aligner}, a novel forced alignment tool that extracts accurate and aligned audio from noisy speech, allowing us to create a high quality kids' data corpus. Collectively, these contributions enhance the capabilities of speech analysis frameworks for children, paving the way for future research and applications in this critical area.

\begin{comment}
Considering the vast size of the dataset, we randomly selected two audio files and transcriptions from the 352 recordings in the ENNI dataset for manual inspection, as reported in Table \ref{table:eval}. Key findings include the following: First, due to noisy transcriptions, MFA \citep{mcauliffe17_interspeech} fails to align the audio segments with the correct transcription.\footnote{This does not imply that MFA is ineffective. MFA performs well with high-quality transcriptions that closely match the audio. However, when the audio/transcription mismatch occurs before alignment, MFA cannot produce meaningful results.} Specifically, missing transcriptions at the start of both audio files lead to a 99.93\% AW Error. MFA attempts to align all words from the beginning, but the absence of corresponding transcriptions causes the alignment process to fail. Second, \textsc{FASA} incorrectly aligns one utterance, missing the "so the" sound at the end, likely due to the speaker’s stutter, which the model does not capture. Lastly, \textsc{FASA}'s performance demonstrates significant improvements over human annotators. As noted by \citet{attia2023kidwhisper}, 5 out of 393 hours of speech in the MyST dataset \citep{pradhan2023science} exhibit errors with WER$>50\%$, contributing to a 3\% increase in WER for the entire training set. In contrast, \textsc{FASA} achieves WER that is an order of magnitude lower (\textbf{13.6$\times$}) without human intervention.
\end{comment}

\begin{table}[!t]
\centering

    \caption{Manual inspections on the generation quality of \textsc{FASA} on two randomly selected audio files and their transcriptions. \textbf{AU:} Aligned Utterances, \textbf{AW:} Aligned Words.}
    \label{table:eval}
\begin{tabular}{ccccc}
\toprule
\textbf{Model} & \textbf{AU} & \textbf{AU Error (\%)}  & \textbf{(AW)}  & \textbf{AW Error (\%)} \\ 
\midrule
\rowcolor{LightGray}MFA & 17 & 16 (94.12\%)   & 1524 & 1523  (99.93\%)   \\ 
\rowcolor{LightCyan}\textsc{FASA} & 81  & 1 (\textbf{1.23\%})   & 903 & 2  (\textbf{0.22\%})    \\
\bottomrule
\end{tabular}
\end{table}
\newpage
\bibliography{iclr2025_conference}
\bibliographystyle{iclr2025_conference}

\newpage
\appendix
\section{Appendix}

\begin{table}[htbp]
\centering
\begin{tabular}{l p{12cm}}
\hline
\hline\\
\rowcolor{LightCyan} 9y old Girl: & \textit{"The frog jumped off the box."} \\
\rowcolor{LLightGray}User: & Determine the gender of the speaker in this clip. \\
\rowcolor{LightRed}\textsc{KidSpeak:}  & The child in the clip is a girl. \\\\
\hline\\
\rowcolor{LightCyan} 4y old Boy: & \textit{"The elephant grabs it and drops it in the water."} \\
\rowcolor{LLightGray}User:  &Transcribe the spoken words into text. \\
\rowcolor{LightRed}\textsc{KidSpeak:}  & The elephant grabs it and drops it in the water. \\\\
\hline\\
\rowcolor{LightCyan} 7y old Boy: & \textit{"Just turn the tap on for water."} \\
\rowcolor{LLightGray}User:  & Evaluate whether the speech in this clip is typical or impaired. \\
\rowcolor{LightRed}\textsc{KidSpeak:} &  The manner of speaking seems natural. \\\\
\hline\\
\rowcolor{LightCyan} 10y old Girl: & \textit{"The hippo jumps in the pool."} \\
\rowcolor{LLightGray}User:  & Estimate the age-group of the speaker from this clip. \\
\rowcolor{LightRed}\textsc{KidSpeak:}  & It is estimated that the speaker may be between 6 - 13 years old. \\\\
\hline\\
\end{tabular}
\caption{Conversation with \textsc{KidSpeak}.}
\end{table}

\subsection{Related Work} \label{sec: lit}

\textbf{Audio-Visual Understanding:}
Contemporary work has attained remarkable feats with regards to understanding the visual and lingual modalities, often in synergy with each other, owing to the large scale availability of datasets facilitating research. Several works including but not limited to LLaVA \citep{liu2024visual}, PaLME \citep{driess2023palm}, Flamingo \citep{alayrac2022flamingo}, BLIP \citep{li2022blip} and GPT-4 \citep{achiam2023gpt} incorporate instruction based tuning in order to incorporate multi-modal capabilities in language models. The training routines for these methods involve tokenization of the image modality allowing for the formation of a sequence which one may incorporate into instruction/text token sequences using self attention (LLaVA, PaLME), cross attention (Flamingo) or through a separate network (BLIP). This often consists of multiple stages wherein the initial set of stages are aimed at priming the learnable parameters for the newer modality. In view of their tremendous potential, we also witness applications of these advancements helping significantly enhance human-AI interaction by improving search engines, supporting creative tasks, and, importantly, advancing accessibility, particularly for the image modality. For instance, these technologies can be utilized to enhance accessibility tools, such as through augmented communication \citep{chanjaradwichai2019trackable}, as assistive learning tools \citep{padmanabha2024voicepilot, kazemitabaar2024codeaid} and sign language recognition systems \citep{gong2024llms}. 

\textbf{Speech based LLMs and Spoken Language Understanding:}
Encoding speech using LLMs is faced with challenges associated with encoding very large sequences of aural representations, given that a 16kHz sampling of one second of audio contains 16000 unique representations of the medium in the frequency domain. However, several recent contributions quantized representations of speech provided by the HuBERT \citep{hsu2021hubert}, as leveraged by the works of Generative Speech Language Modeling \citep{lakhotia2021generative}, TWIST \citep{hassid2024textually} and SpeechGPT \citep{zhang2023speechgpt} wherein based on the applications, the representations are used in transformer \citep{vaswani2017attention} based encoder decoder systems or are combined with the textual embeddings in order to construct an interactive multi-modal system using instruction based tuning, with generative capabilities in audio. Another encoding scheme for audio is the use of log-mel spectrograms, processed using encoders such as \textsc{Whisper} \citep{radford2023robust}, \textsc{Wav2Vec} \cite{baevski2020wav2vec20frameworkselfsupervised}, Conformer \citep{gulati2020conformer} and the AST \citep{gong2021ast}, in order to generate discrete representations which may similarly be combined with textual representations for understanding and generative tasks. \cite{fathullah2024prompting} incorporate the conformer architecture in order to enable speech recognition in LLMs. Spectron \citep{nachmani2023spoken} uses the conformer encoder and splits the aural representations into prompt and continuation, enabling the continuation of aural speech and language using prompt during inference. \cite{zhao2023librisqa} use the WavLM \citep{chen2022wavlm} and the \textsc{Wav2Vec} encoders whereas \cite{gong2023listen, ghosh2024gamalargeaudiolanguagemodel} utilize the AST model and develop a question answering based training procedure to inculcate understanding in a LLM using the aural representations. 

\textbf{Kids' Speech:}
A predominant drawback of the concurrent works is the lack of support for nuanced speech based on a variety of accents, dialects and intonations, or developing or disordered speech as often characterized by kids. This challenge is further exacerbated by the relative lack of attention and research in the field of children's speech recognition and correction, resulting in a limited number of robust solutions tailored specifically to these needs. \cite{liao2015large} use a LSTM \citep{10.1162/neco.1997.9.8.1735} and CLDNN \citep{sainath2015convolutional} based architectures towards transcription and reduction of offensive generations. \cite{plantinga2019towards} use a GRU \citep{cho2014learning} based architecture with alignment loss to discourage generation during silence and a teacher student loss in order to improve transcription performance. \cite{ramesh2022beach} leverage a masked word prediction based cloze task inspired by BERT based encoder systems in order to correct offensively transcribed speech by existing ASR systems. \citep{venkatasubramaniam2023end} develop an LSTM based disfluency detection and classification architecture over an existing ASR system in order to enhance transcription. However, to the best of our knowledge, this is the first work that proposes utilizing a large language model (LLM) as a multi-task model with diagnostic capabilities within the speech domain. \textit{We anticipate that this represents a promising avenue for future research, offering substantial benefits in the realm of speech therapy for kids and complementing the extensive efforts of Speech-Language Pathologists (SLPs)}.

\textbf{Forced-Alignment Toolkits:} \label{forced_alignment_related_works} Traditionally yet still prevalently, alignment between the audio and its transcription is done via human annotators on various software \citep{praat,grover2020audino}. However, as discussed earlier, such practice is not scalable for large datasets. \citet{KISLER2017326} contains some parts of a complete forced-alignment pipeline, but it does not address the fundamental problem of aligning audio with its transcription.
\citet{GAN2019} uses generative-adversarial networks (GAN) to perform data augmentation on children ASR dataset, but their work does not introduce diverse \textit{new} data to the field. Recently, the Talkbank project announced its data processing pipeline that converts raw audio into CLAN-annotated transcriptions \citep{Liu_MacWhinney_Fromm_Lanzi_2023}. While their work uses a similar backbone structure as ours, their complete pipeline relies on transcription generated by ASR models, whereas we faithfully adhere to the provided transcription as the ground truth. Thus, on downstream tasks such as fine-tuning ASR models, our dataset will be more usable because datasets generated by ASR models might cause severe degradation according to \citep{radford2022whisper}. On the other hand, there have been several works on forced-alignment ASR datasets with the assistance of human-labeled transcriptions \citep{mcauliffe17_interspeech, POnSS, zhang2023speechgpt,Liu_MacWhinney_Fromm_Lanzi_2023}, with Montreal-Forced-Aligner (MFA) being the most popular toolkit \citep{mcauliffe17_interspeech}. MFA incorporates Kaldi \citep{Povey_ASRU2011} as the backbone, which uses the Gaussian Mixture Model (GMM) for its transcription generation process. However, while MFA works well with carefully annotated transcriptions, it requires the transcription to have a perfect match with the audio. That is, $A_1 \to \{T_1,T_2,...,T_i\}$, $A_2\to \{T_{i+1},T_{i+2},...,T_{j}\}$, and so forth. \citet{Liu_MacWhinney_Fromm_Lanzi_2023} faces similar issues that it lacks global matching ability between audio and transcription, hindering its usage in some subsets of the speech corpus. %Thus, under the two practical assumptions defined in Section 2.1, %\cite{POnSS} requires extensive human label, whereas 
%MFA \cite{mcauliffe17_interspeech} does not generate satisfactory results. 
Moreover, while recent multi-modal large language models (MLLM) might have the potential of automating the alignment process \citep{zhang2023speechgpt}, they are much more resource-intensive compared to specific ASR models.

\subsection{Phonetically Endowed Multihead Whisper}
\subsubsection{Decoder Alignment in Multi-head \textsc{Whisper}} \label{sec: mh_justification}
The two alignment mechanisms address a critical need for alignment between phonetic and orthographic transcriptions within multi-modal speech processing systems. This alignment is essential for several reasons:

\begin{itemize}
\item \textit{Phonetic and Orthographic Consistency}:
   Phonetic transcriptions represent the pronunciation of words, focusing on sounds, while orthographic transcriptions represent the written form of language. Aligning these two modalities ensures that the pronunciation (phonetics) and spelling (orthographics) are consistent with each other. This consistency is crucial for tasks such as speech recognition and language learning, where accurate mapping between spoken and written forms is required.

\item \textit{Enhanced Encoder Utility}:
   By aligning phonetic and orthographic transcriptions through the shared encoder, the model benefits from enriched feature representations. The encoder, which is common to both decoders, learns to produce more comprehensive and phonetically aware representations. This shared learning helps the encoder capture nuances that prove to be critical for improved understanding of pronunciation based nuances necessary for improved diagnostic capacities.

\item \textit{Robust Multi-Modal Learning}:
   Aligning the hidden states of phonetic and orthographic decoders allows the system to leverage complementary information from both transcriptions. Phonetics provides insights into pronunciation nuances, while orthographics offer context about spelling and grammar. The combined insights from both modalities lead to a more robust and versatile model capable of handling diverse linguistic tasks.
\end{itemize}

In the following, we demonstrate the phonetic capacities of our scheme using comparisons with \textsc{Whisper} and \textsc{Wav2Vec} trained over \textsc{TIMIT}. The phonetic alphabet used natively by \textsc{TIMIT} is illustrated here for selected samples. We notice that all of the models capture a meaningful pronunciation for each word of the examples listed. However, the targeted alignment scheme of our method captures the nuances in pronunciation, audible in the ground truth phonetic captioning, enabling a more accurate transcription and henceforth, better encoder representations, for therapeutic downstream tasks.

\begin{tcolorbox}[title=1. SCRIPT: She had your dark suit in greasy wash water all year, colback=gray!5!white, colframe=gray!75!black]
     \textbf{Groundtruth:} shih hhehjh jhih \textbf{pau} dahk suw n \textbf{pau} grishih waash waadxer aal yiher \\
     \textbf{Wav2Vec:} shix \textcolor{red}{hvehjh jhuh} \textbf{pau} dahk \textcolor{red}{suxq en \textbf{pau}} gri\textcolor{red}{six wao}sh \textcolor{red}{waodxax aol} yih\textcolor{red}{axr} \\
     \textbf{Whisper:} shi\textcolor{red}{x hvehjh} jhih \textbf{pau} dahk \textcolor{red}{suxq en} \textbf{pau} gris\textcolor{red}{xix} wa\textcolor{red}{osh waodxaxr aol} yih\textcolor{red}{axr} \\
     \textbf{Whisper ours:} shih hhehjh jhih \textbf{pau} dahk suw n \textbf{pau} grishih waash waadxer aal yiher
\end{tcolorbox}

% Second box for the second script
\begin{tcolorbox}[title=2. SCRIPT: Don’t ask me to carry an oily rag like that, colback=gray!5!white, colframe=gray!75!black]
    \textbf{Groundtruth:} down aes \textbf{pau} my th \textbf{pau} kehriy ihn oylih raeg lay dhae \\
    \textbf{Wav2Vec:} down aes \textbf{pau} my \textcolor{red}{\textbf{pau} tx} \textbf{pau} kehriy \textcolor{red}{ixn qoyliy} raeg lay dhae\\
    \textbf{Whisper:} down aes \textbf{pau} my \textcolor{red}{\textbf{pau} tx} \textbf{pau} kehriy \textcolor{red}{ixn qoy}lih raeg lay dhae\\
    \textbf{Whisper ours:} down aes \textbf{pau} my th \textbf{pau} kehriy ihn oylih raeg lay dhae
\end{tcolorbox}

\subsubsection{Disorder detection}
In the following, we demonstrate the capacity of the phonetically endowed \textsc{Whisper} model using speech from the Ultraphonix dataset \cite{eshky2019ultrasuite}. The \textsc{Whisper} model was trained using the Multihead alignment scheme described in Section \ref{sec: mh_method} using the \textsc{TIMIT} corpus \citep{garofolo1993timit}. Subsequently, we conduct inference using the phonetic decoder of the model over the speech of a 4 year old boy undergoing therapy for phonological disorder. The child mistakenly uses the sound of \textit{"da"} in place of \textit{"ga"} for words. For instance, the pronunciation for the word "luggage" (\textbf{"LUG-ij"}) and "gore" (\textbf{"GOw-ar"}) here are made as \textit{"LAD-ij"} and \textit{"DOw-ar"}. However, post therapy, the child learns to pronounce clearly as is captured by our model. We use the \textsc{TIMIT} phonetic transcription code here. \textbf{pau} indicates a pause in speech.

\begin{tcolorbox}[colback=gray!5!white, colframe=black!75!black,]
\textbf{SCRIPT:} \textcolor{red}{gore} \textcolor{orange}{gate} \textcolor{blue}{get} \textcolor{green}{luggage} \newline

\textbf{Instructor:}  \textcolor{red}{g ow axr} \textbf{pau} \textcolor{orange}{g ey t} \textbf{pau} \textcolor{blue}{g eh t} \textbf{pau} \textcolor{green}{l ah g ux jh}  \newline
\textbf{Child Pre:}  \textcolor{red}{d ow ax} \textbf{pau} \textcolor{orange}{d iy t} \textbf{pau} \textcolor{blue}{d ae t} \textbf{pau} \textcolor{green}{l ah d ix jh}  \newline
\textbf{Child Post:}  \textcolor{red}{g ow aa } \textbf{pau} \textcolor{orange}{g ih g eh ix t} \textbf{pau} \textcolor{blue}{g ae t} \textbf{pau} \textcolor{green}{l ah g ih jh} 
\end{tcolorbox}

As is evinced in the illustration, the phonetically endowed \textsc{Whisper} correctly detects the improvements in pronunciation in \textbf{pre-} vs \textbf{post-} therapy of the child, thereby allowing for tailored features for targeted therapy based downstream tasks such as those implemented in \textsc{KidSpeak}.

\subsection{Features of \textsc{FASA}} \label{sec: fasa_features}
Similar to existing auto-alignment toolkits, \textsc{FASA} requires an audio file and its corresponding transcription. However, due to high uncertainty in the raw dataset, \textsc{FASA} assumes only a minimal input format and does not require the transcription to be accurate. The ground truth (GT) for an utterance \( A_k \) is defined by Equation \ref{eq} in the \textsc{FASA} pipeline. In contrast to previous forced-alignment toolkits, \textsc{FASA} deliberately ignores utterances without valid transcriptions, thereby enhancing quality.

\textsc{FASA} also incorporates beneficial design elements from established toolkits to enhance user convenience, following the same design principles as \textsc{MFA}. Users need only to place the audio file and its transcriptions in a designated folder before executing the program, after which all processes are fully automated, enhancing the user experience. \textsc{FASA} allows users to select and manually input transcriptions for utterances when the provided transcriptions are suspected to be inaccurate, ensuring precision and user control. Additionally, \textsc{FASA} features an optional post-generation check to automatically exclude incorrect alignments, minimizing errors from the underlying model. 

\subsection{Datasets} \label{sec: datasets}
The Core-Ultraphonix (UPX) subset of the Ultrasuite repository \citep{eshky2019ultrasuite} provided the labels for speech with pathologies. Additionally, we incorporate the Children's Speech Recording (CSR) dataset by \cite{kennedy2017child}. With \textsc{FASA}, we convert subsets of the Child Language Data Exchange System (CHILDES) \citep{CHILDES} that contain English children's speech. Specifically, we use a collection of 352 children from ages 4 to 9. The children are performing the Edmonton Narrative Norms Instrument (ENNI) test \citep{ENNI}. To the best of our knowledge, this generated dataset will be the first at-scale high-quality dataset for young children from clinical recordings that is fully compatible with modern DL systems. While we only use the subset from CHILDES with rich clinical information for \textsc{KidSpeak}, \textsc{FASA} remains a generic forced-alignment toolkit that can extract many more datasets than the one used.
%\dcheng{can delete the previous sentence and Appendix A.7 if you use three datasets.}

\subsection{Speech Disorder Classes} \label{sec: disorders_defn}

The following provides a broad explanation of various speech-related disorders, along with seminal and intriguing citations in the field of speech-language pathology that \textsc{KidSpeak} is capable of diagnosing based on the speech patterns exhibited by the child.

\begin{itemize}
    \item \textit{Inconsistent Phonological Disorder}: This pediatric speech sound disorder is characterized by the inconsistent production of the same words across repeated trials \citep{dodd2024diagnosing}. For example, a child may say \textit{"bat,"} \textit{"gat,"} and \textit{"at"} instead of \textit{"cat,"} or produce \textit{"log"} for \textit{"dog"} one day and \textit{"fog"} the next. Moreover, a child may say \textit{"fider,"} \textit{"sider,"} and \textit{"pider"} when attempting to pronounce \textit{"spider"} \citep{dodd2010language, carter2019etiology}.

    \item \textit{Consistent Phonological Disorder}: In contrast, this disorder is marked by the child's ability to produce the same errors consistently when attempting to articulate the same word. For instance, a child may reliably say \textit{"tup"} instead of \textit{"cup"} or \textit{"wabbit"} for \textit{"rabbit."} Such patterns indicate a stable phonological processing issue, as the child consistently makes the same substitutions or distortions \citep{felsenfeld1995familial, bleile2002evaluating}.

    \item \textit{Phonological Delay}: This specific speech sound disorder entails developmental phonological errors that align with typical speech development patterns but persist longer than expected, often for six months or more, which can impact clarity and sound production \citep{orsolini2001nature}. Children acquire speech by learning entire words rather than individual sounds; as their speech matures, they categorize words by their components, often simplifying sounds or sequences into easier alternatives (e.g., saying \textit{"ca"} for \textit{"cat"}) \citep{waring2022differentiating}.

    \item \textit{Vowel Disorder}: Vowel disorders are marked by difficulties in the positioning and sequencing of the articulators, particularly the tongue and lips, affecting vowel quality and accuracy. Incorrect positioning can lead to issues with vowel production, such as excessively long vowels or distortions. For instance, vowels may be partially voiced due to challenges in controlling vocal fold vibration, or they may exhibit excessive nasality from difficulties managing velopharyngeal closure. These spatial, temporal, and coordination difficulties often result in challenges in vowel production \citep{gibbon2002therapy, ball2002vowel}. Additionally, children may struggle with vowel lengthening or shortening, such as elongating the vowel in \textit{"see"} for \textit{"sit"} or shortening it in \textit{"cat"} as \textit{"kit,"} with omissions occurring as well (e.g., saying \textit{"bll"} instead of \textit{"ball"}) \citep{stoel2008vowel}.

    \item \textit{Articulation Disorder}: This type of speech sound disorder is characterized by difficulties in accurately producing speech sounds due to the imprecise use of the lips, tongue, or throat. Individuals may demonstrate various symptoms, including the omission of sounds (e.g., final consonants), distortion of sounds (e.g., producing an \textit{"s"} sound with a whistle), and challenges in coordinating the movements of their lips, tongue, teeth, palate, and lungs \citep{hall1978follow, rvachew1989perception}.

    \item \textit{Childhood Apraxia of Speech}: Childhood apraxia of speech (CAS) is a neurological speech sound disorder characterized by impaired precision and consistency of movements underlying speech, absent neuromuscular deficits (e.g., abnormal reflexes or tone) \citep{davis1998developmental, american2007childhood, kummer2007prevalence}. Children with CAS may encounter difficulties in speech production, such as trouble transitioning smoothly between sounds and syllables, groping movements of the jaw, lips, or tongue, vowel distortions, incorrect stress patterns (e.g., pronouncing \textit{"banana"} as \textit{"BUH-nan-uh,"}) equal emphasis on all syllables (e.g., saying \textit{"BUH-NAN-UH,"}) separation of syllables with pauses, inconsistency in errors when repeating words, and voicing errors (e.g., saying \textit{"down"} instead of \textit{"town"}) \citep{carter2019etiology}.
\end{itemize}

\subsection{FASA workflow pseudocode}
\label{fasa_pseudocode}

Here, we provide the pseudocode for the third module of \textsc{FASA}'s workflow in Algorithm \ref{algo:algo}. \textsc{FASA} uses a sliding window algorithm with two thresholds to determine the final two subsets of audio segments. In the algorithm, $DIS$ is the Levenshtein distance between two sentences.

\newcommand{\Input}[1]{\algrenewcommand{\alglinenumber}[1]{Input: \ \setcounter{ALG@line}{\numexpr##1-1}} #1}
\newcommand{\Step}[1]{\algrenewcommand{\alglinenumber}[1]{Step ##1: } #1}
\newcommand{\NoNumber}{\algrenewcommand{\alglinenumber}[1]{\setcounter{ALG@line}{\numexpr##1-1} \ \ \ \ \ \ \ \ \ \ }}
\newcommand{\Output}[1]{\algrenewcommand{\alglinenumber}[1]{Output:\setcounter{ALG@line}{\numexpr##1-1}} #1}
\begin{algorithm*}[!ht]
	\caption{sliding window to find the best matching} 
    \label{algo:algo}
	\begin{algorithmic}[1]
        
         \Input \State $A$, $T=\{T_1,...T_m\}$, $\bar{T}$, alignment threshold $\sigma_a$, inclusion threshold $\sigma_i$.
            \Step \State Initialize holder for dataset of aligned segments: $\text{DATA}_{align}$ = []
            \NoNumber \State Initialize holder for questionable segments:  $\text{DATA}_{verify}$ = []
		\Step \For {$A_k \in A$}
                \NoNumber \State Get $A_k$'s transcription: $\bar{T_{A_k}} = \{\bar{T_{i}}...\bar{T_{j}}\}\ \in \bar{T}$
                \NoNumber \State Initialize minimum distance $D_{min} = \infty$, best starting index $BEST_i$, best length $BEST_l$
			\NoNumber\For {$a=1,2,\ldots,m$}
                        \For {$b=1,2,\ldots,(j-i)$}   
				\If {$DIS(\bar{T_{A_k}}, T[a:a+b+1]) < D_{min}$}
                        \State $D_{min}$ = $DIS(\bar{T_{A_k}}, T[a:a+b+1])$
                        \State $BEST_i$ = $a$
                        \State $BEST_l$ = $b+1$

                    \EndIf
			\EndFor
            \EndFor
            
            \Step \State let $GT_k = T[BEST_i:BEST_i+BEST_l]$
            \NoNumber\If {$WER$ ($GT_k$, $\bar{T_{A_k}}$) $< \sigma_i$}
                \NoNumber\If {$WER$ ($GT_k$, $\bar{T_{A_k}}$) $< \sigma_a$}
                    \NoNumber \State append $(A_k, GT_k)$ to $\text{DATA}_{align}$
                    \Else
                    \NoNumber \State append $(A_k, GT_k, \bar{T_{A_k}} )$ to $\text{DATA}_{verify}$
                \EndIf
            \EndIf
		\NoNumber \EndFor

    \Output \State $\text{DATA}_{align}$, $\text{DATA}_{verify}$
	\end{algorithmic} 
\end{algorithm*}

\subsection{Configuration}
\begin{table}[H]
    \centering
    \caption{Training setup for the methods.}
    \begin{tabular}{lccc}
        \toprule
        \textbf{Attribute} & \textbf{PandaGPT} & \textbf{KidSpeak} & \textbf{MH-Whisper} \\
        \midrule
        Peak learning rate & 5e-5 & 5e-5 & 1e-4  \\
        Batch size & 64 & 64 & 4 \\
        Accumulate Steps & 8 & 8 & --  \\
        Max length & 512 & 512 & 448  \\
        LoRA rank & 16 & 16 & -- \\
        LoRA alpha & 32 & 32 & -- \\
        Training steps & 10000 & 10000 & 63525 \\
        Trainable parameters & 16.78M & 16.78M & 553.1M \\
        Training device & 4xA6000 & 4xA600 & 1xA5000 \\
        \bottomrule
    \end{tabular}\label{tab: config}
\end{table}
    
\begin{comment}
\subsection{FASA extracted datasets}
\label{fasa_extracted}

While we use only the ENNI subset for \textsc{KidSpeak}, \textsc{FASA} is able to generate high-quality datasets from any audio/transcription pair. Here in Table \ref{table:dataset_fasa}, we report other subsets that are generated by \textsc{FASA} from the CHILDES which we do not use due to lack of sufficient speech by children in them, as determined by manual inspection.

\begin{table}[h]
    \centering
    \caption{Specifications for the datasets used by \textsc{KidSpeak}.}
    \label{table:dataset_fasa}
    \small
    \begin{tabular}{cccc}
    \hline
         Dataset &  \# of Speakers & \# of Utterances & Time (H:M:S)  \\ \hline
        %Clinical-Eng & 1540 &59539 & 30:11:40 & overlapping with ENNI\\
        %Clinical-Other & 292& 59539 & 4:18:6 & utterances too short \\
        English-NA & 5332 & 778978  &285:18:22 \\
        \hline
    \end{tabular}
\end{table}
\end{comment}

\end{document}